\documentclass[%
 superscriptaddress,
 amsmath,amssymb,
 prb,
 twocolumn,
]{revtex4-2}

\usepackage{graphicx}
\usepackage{dcolumn}
\usepackage{bm}

\usepackage[table,svgnames]{xcolor} 
\usepackage{float}

\usepackage{color}
\usepackage[normalem]{ulem}
\usepackage{hyperref}

\definecolor{eggplant}{RGB}{180,33,147}

\definecolor{grey}{rgb}{.6,.6,.6}

\begin{document}

\title{Tensorized orbitals for computational chemistry}

\author{Nicolas Jolly}
\affiliation{Univ. Grenoble Alpes, CEA, Grenoble INP, IRIG, PHELIQS, 38000 Grenoble, France}
\author{Yuriel N\'u\~nez Fern\'andez}
\affiliation{Universit\'e  Grenoble Alpes, CNRS, Institut N\'eel, 38000 Grenoble, France}
\author{Xavier Waintal}
\affiliation{Univ. Grenoble Alpes, CEA, Grenoble INP, IRIG, PHELIQS, 38000 Grenoble, France}
\email{xavier.waintal@cea.fr}

\date{\today} 

\begin{abstract}
Choosing a basis set is the first step of a quantum chemistry calculation and it sets its   maximum accuracy. 
This choice of orbitals is limited by strong technical constraints as one must be able to compute a large number of six dimensional Coulomb integrals from these orbitals. 
Here we use tensor network techniques to construct representations of orbitals that essentially lift these technical constraints. We show that a large class of orbitals can be put into ``tensorized'' form including the Gaussian orbitals, Slater orbitals, linear combination thereof as well as new orbitals beyond the above. Our method provides a path for building more accurate and more compact basis sets beyond what has been accessible with previous technology. 
As an illustration, we construct optimized tensorized orbitals and obtain a 85\% reduction of the error on the energy of the $H_2$ molecules with respect to a reference double zeta calculation (cc-pvDz) of the same size.
\end{abstract}

\maketitle

\section{Introduction}
The very first step of a first principles many-body calculation is to discretize the problem onto {\it some} finite basis set. On the quality of this discretization depends the maximum accuracy that may be reached in the calculation and therefore its usefulness for making predictions \cite{woon1995,peterson2012,cances2017}. The commonly accepted target of $1.6 mHa$
(around $500 K$) for ``chemical accuracy'' typically requires large basis sets to 
approach the continuum sufficiently well. There exists an immense body of literature devoted to the construction of optimized basis sets for a wide variety of situations. Among others, those include plane waves, wavelets \cite{white2017},  Wannier function \cite{pizzi2020}, Slater \cite{forster2021} and many flavours of Gaussian orbitals \cite{woon1995,pople2003,jensen2013,zhou2021}. Extrapolating the results to the Complete Basis Set (CBS) limit is still an important difficulty in practice 
\cite{bruneval2021,forster2021}.

In the context of chemistry, the Gaussian orbitals overwhelmingly dominate
the literature. Their popularity stems from two important properties. The first is (P1) their compactness, i.e. a few Gaussians centered on the different nuclei are sufficient to give a reasonably accurate approximation of the atomic orbitals. Second, and perhaps more importantly (P2), the $6$-dimensional integrals appearing in the calculation of the electron-electron interaction matrix elements can be computed analytically for Gaussian orbitals. Property (P2) sounds somewhat technical.
Yet, it is the crucial point that has allowed Gaussian orbitals to thrive and become ubiquitous in computational chemistry since these matrix elements are the starting points on which all further calculations are based. 
Despite these favourable aspects, Gaussian orbitals do have a number of important limitations: the convergence to the CBS limit is slow, core electrons are poorly described (in part because Gaussians lack the cusp that true atomic orbitals have close to the nuclei core; also the tail of the orbitals is in general not Gaussian) and so are delocalized orbitals above the ionization threshold. 

\begin{figure*}[t!]
    \centering
    \includegraphics[width=\textwidth]{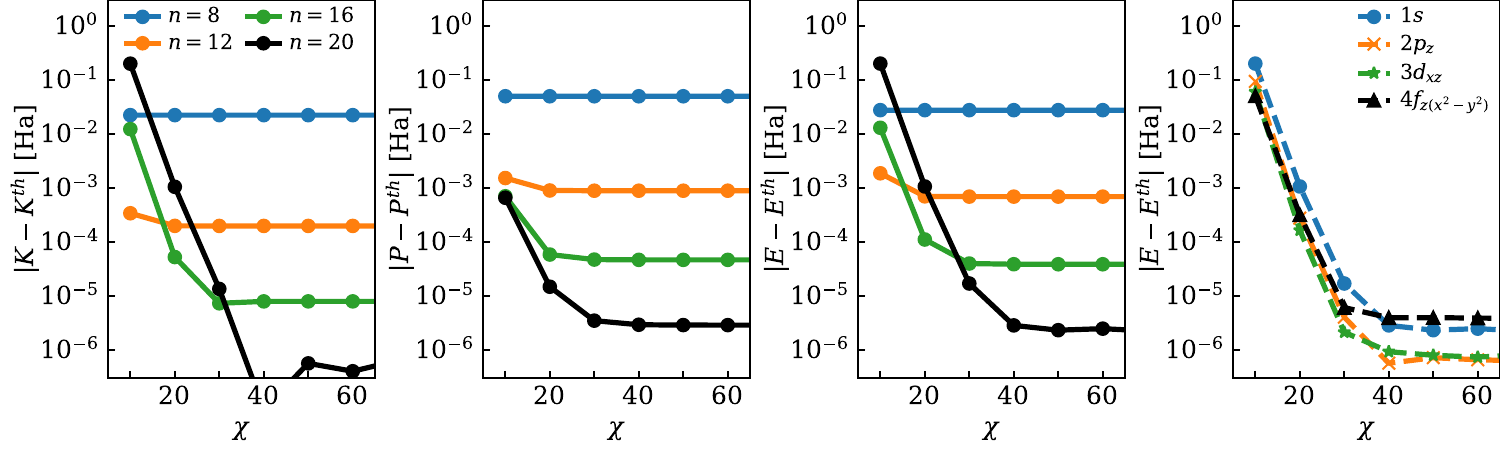}
    \caption{Tensorisation of Slater orbitals. Error versus bond dimension $\chi$ for the energy of the exact $1s$ orbital of the hydrogen atom (first three panels) and other orbitals ($1s$, $2p_z$, $3d_{xz}$ and $4f_{z(x^2-y^2)}$, last panel). First three panels: error on the kinetic energy $K$, nuclei potential energy $P$ and total energy $E=K+P$ for different grid discretization $n=8,12,16$ and $20$.  Last panel: $n=20$ except for the $4f$ orbital for which $n=22$. Energies in Hartree.
    \label{fig:1}
    }
\end{figure*}

In this letter, we do {\it not} propose another set of orbitals. Rather, we propose a novel {\it representation} of orbitals, using tensor networks \cite{schollwoeck2010,orus2014,orus2019,Bridgeman2017}. This representation contains a large class of orbitals that includes Gaussian orbitals, Slater orbitals, plane waves, new types calculated directly in real space as well as any combinations (e.g. molecular and natural orbitals).
At the same time, it naturally solves the problem of computing the needed matrix elements (P2): once ``tensorized'', the orbitals provide fast and robust access to all the usual mathematical objects needed by chemistry packages to proceed with the calculation. 

\section{Problem formulation} A molecule with $N$ electrons is described by the positions $\vec R_\alpha$ and atomic number $Z_\alpha$ of its nuclei. We consider a basis set of $M$ orbitals $\phi_i(\vec r)$. Within this basis set, the many-body problem that one needs to solve has the following Hamiltonian,
\begin{equation}
H = \sum_{ij\sigma} H_{ij} c^\dagger_{i\sigma} c_{j\sigma} + 
\sum_{ijkl\sigma\sigma'} V_{ijkl} c^\dagger_{i\sigma} c^\dagger_{j\sigma'} c_{k\sigma'} c_{l\sigma} 
\end{equation}
where $c^\dagger_{i\sigma}$ ($c_{i\sigma}$) creates (destroys) an electron in orbital $i$ and spin $\sigma$. They satisfy the anti-commutation rule
$\{c^\dagger_{i\sigma} , c_{j\sigma'} \} = \delta_{\sigma\sigma'} S_{ij}$ where
$S_{ij}$ is the overlap matrix. The problem is therefore defined by the objects $S_{ij}$, $H_{ij}$ and $V_{ijkl}$ whose expression in terms of the orbitals
$\phi_i(\vec r)$ is,
\begin{eqnarray}
\label{eq:S}
S_{ij} &=& \int d\vec r \ \phi_i(\vec r) \phi_j(\vec r) \\
\label{eq:H}
H_{ij} &=& K_{ij} + P_{ij} \nonumber \\
K_{ij} &=& \int d\vec r \ \phi_i(\vec r) 
\left[ \frac{-\hbar^2}{2m}\Delta \right] \phi_j(\vec r) \nonumber \\
P_{ij}&=& -\int d\vec r \ \phi_i(\vec r) 
\left[\sum_\alpha \frac{Z_\alpha e^2}{4\pi\epsilon |\vec r - \vec R_\alpha|} \right] \phi_j(\vec r) \\
\label{eq:V}
V_{ijkl} &=& e^2\int d\vec r_1 d\vec r_2 \ 
\frac{\phi_i(\vec r_1)\phi_j(\vec r_2)\phi_k(\vec r_2)\phi_l(\vec r_1)}{4\pi\epsilon |\vec r_1 - \vec r_2|}
\end{eqnarray}
For a set of orbitals to be useful, one must be in position to compute Eqs. \eqref{eq:S}, \eqref{eq:H} and \eqref{eq:V} quickly and accurately. In particular the bottleneck of this calculation is given by the interaction matrix elements Eq. \eqref{eq:V} which contains a large number $\propto M^4$ of $6$-dimensional integrals.

\section{Tensorized orbitals} 
\subsection{Quantics representation of an orbital} Let's consider an orbital $\phi(\vec r)$ inside the cube $\vec r = (x,y,z) \in [0,b]^3$. The first step of the construction of the tensorized orbitals is to discretize the segment $[0,b]$ onto an {\it exponentially} dense grid of $2^n$ equally spaced points. Each of these points is labeled by $n$ bits $x_1x_2....x_n \in \{0,1\}^n$ such that
\begin{equation}
\frac{x}{b} = \frac{x_1}{2}+\frac{x_2}{2^2}+\dots +\frac{x_n}{2^n}
\label{eq:quantics}
\end{equation}
with similar equations for the $y$ and $z$ coordinates. The discretization of $\phi(\vec r)$ on this grid is a very large tensor $\Phi_{\vec r}$ with $3n$ indices indexing the $(2^n)^3$ different values of the orbital on the grid,
\begin{equation}
\Phi_{\vec r} \equiv \Phi_{x_1x_2...x_ny_1y_2...y_nz_1z_2...z_n}
\end{equation}
Second, we represent the tensor $\Phi_{\vec r}$ as a tensor train (also known as matrix product state or MPS) in the so-called \emph{grouped ordering} (other orderings are discussed in the Appendix \ref{bit_ordering}),
\begin{equation}
\label{eq:mps_grouped}
\Phi_{\vec r} = \prod_{a=1}^{n} M_{a}(x_a) \prod_{a=1}^{n} M_{a+n}(y_a) \prod_{a=1}^{n} M_{a+2n}(z_a).
\end{equation}
The tensor $\Phi_{\vec r}$ is now represented in terms of $2\times 3n$ matrices $M_p$ of size (up to)
$\chi\times\chi$. This is the quantics representation \cite{oseledets2009,khoromskij2011,nunez2024} of the orbital. Such a decomposition is always possible but the bond dimension $\chi$ may be exponentially large $\chi \sim 2^{3n/2}$. The magic of tensor trains is that for many mathematical objects the convergence of the tensor train with $\chi$ is very fast and very accurate results can be achieved with small values of $\chi$ ($\chi < 100$ in this work). In other words, many orbitals can be \emph{compressed} using tensor trains.
MPS plays a central role in computational many-body theory \cite{schollwoeck2010,orus2014,orus2019}, in particular in the context of the density matrix renormalization algorithm (DMRG) \cite{white1992}. The joined usage of MPS with the quantics representation is much more recent and has only recently started to be explored \cite{lubasch2018,ripoll2021,ye2022,gourianov2022,shinaoka2022,ritter2023}. In the rest of this letter, we propose three different algorithms to arrive at the tensorized representation of an orbital Eqs. \eqref{eq:mps_grouped}. We will further discuss how, once the tensorized representation has been obtained, one can proceed with the calculation of $S_{ij}$, $H_{ij}$ and $V_{ijkl}$ thereby closing the gap to continue the calculation further with any many-body technique (e.g. Hartree-Fock, DMRG, coupled clusters...).

 The central algorithm that makes the present work possible is the recently introduced
 Tensor Cross Interpolation (TCI) learning algorithm \cite{oseledets2011,dolgov2020} that allows one to construct the tensor train from a few $\sim n \chi^2$ calls to the function $\phi(\vec r)$. We rely on the \emph{xfac} implementation and extensions of TCI described in detailed in \cite{nunez2024, nunez2022} to which we also refer for an introduction to the algorithm. For the needs of this letter, it is sufficient to know that the input of TCI is a tensor (in the form of a function that can be called for any values of the indices) and its output is the tensor train Eqs. \eqref{eq:mps_grouped} together with an estimate of the error (systematically controled by increasing the bond dimension $\chi$).

\subsection{First example: tensorization of the orbitals of the hydrogen atom.}

As a first illustration, we construct the tensorized representation of the exact Slater orbitals of the hydrogen atom using TCI. The $1s$ orbital is extremely simple, $\phi(\vec r) \propto \exp (-\sqrt{x^2+y^2+z^2}/a_0)$ ($a_0$: bohr radius). Yet, computing precisely the different matrix elements for such orbitals centered around different nuclei positions is highly non-trivial \cite{velde2001,forster2021} which has restricted the use of Slater orbitals so far despite their potential benefits  (correct cusps and tails). The first three panels of Fig. \ref{fig:1} show the error respectively for the kinetic energy $K$, potential energy $P$ and total energy $E=K+P$ for the ground state of the hydrogen atom (the algorithms for these calculations will be detailed below). We observe that a very moderate value of $\chi$ is needed to achieve very precise accuracy $<0.01m{\rm Ha}$. On the other hand, the calculations of $K$ and $P$ requires a very fine grid with $n\ge 16$. This is not a issue with the quantics representation since the computational cost is linear in $n$. Indeed, the calculations with $n=20$ (black curve) is only marginally more difficult than the calculation for $n=12$ (orange) despite the fact that, formally, $n=20$ corresponds to a grid that contains an astronomically large $\sim 10^{18}$ number of points ($\sim 10^{36}$ points for the $6$-dimensional integrals). In the last panel of Fig. \ref{fig:1}, we show the error for $p$, $d$ and $f$ orbitals. One finds that the accuracy depends only very weakly on the actual shape of the orbital. For the $4f$ orbital, we had to increase $n$ up to $n=22$ since this orbital is much more extended than $1s$. We found that TCI algorithm works equally well for all the orbitals that we have tested; they only need to be known explicitly, i.e.
one can compute $\phi(\vec r)$ for any value of $\vec r$. This includes in particular any combination of Gaussians as we have checked explicitly.
 
 \begin{figure*}[t!]
    \centering
    \includegraphics[width=\textwidth]{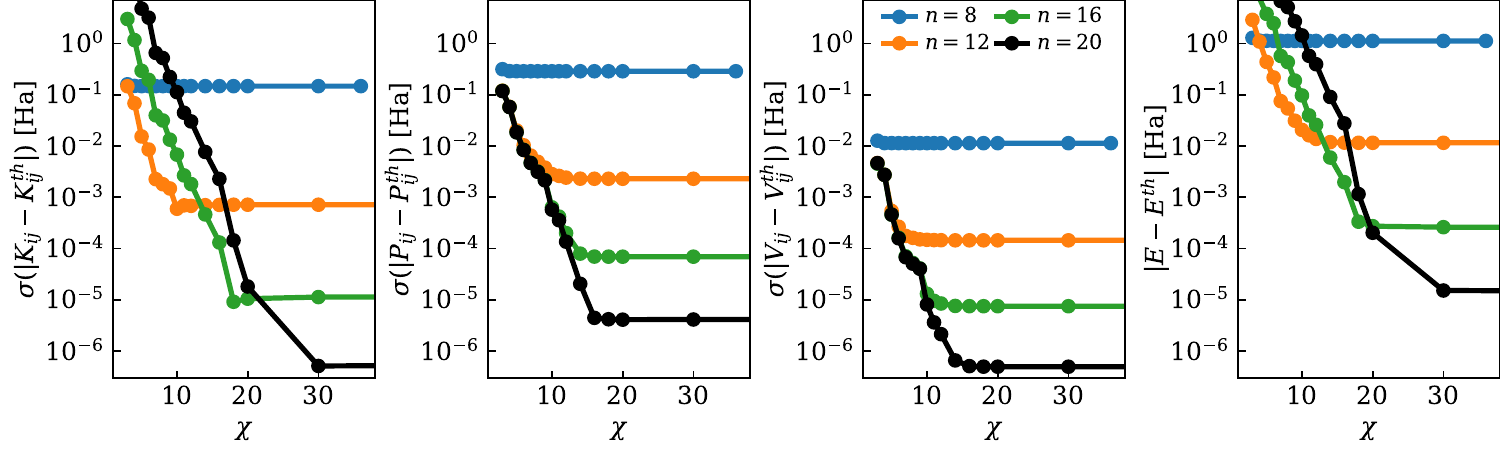}
    \caption{Tensorisation of Gaussian orbitals: LiH molecule in the STO-6G basis set.
    Error versus bond dimension $\chi$ of the orbitals for the different matrix elements (first three panels, $\sigma$ is the standard deviation) and overall CCSD(T) energy (last panel) for different discretization parameters $n=8,12,16$ and $20$. Distance between the nuclei: $d_{Li-H} = 2.8571 a_0$. 
    Reference calculation: pyscf package \cite{sun2018}.
    \label{fig:2}
    }
\end{figure*}
\subsection{Algorithms for the computation of $S_{ij}$, $H_{ij}$ and $V_{ijkl}$} The crucial property
of tensorized orbitals is the ability to compute the interaction matrix elements needed for subsequent many-body calculations at an exponentially reduced $O(n)$ cost. The simplest one is $S_{ij}$. Indeed, in the framework of MPS, $S_{ij}=V_0 \sum_{\vec r}\Phi^i_{\vec r}\Phi^j_{\vec r}$ is simply the scalar product between two MPS. Here $V_0\equiv b^3/2^{3n}$ is the elementary grid cell volume. The algorithm for contracting this tensor network belongs to the standard toolbox of MPS \cite{schollwoeck2010} and has a mild complexity $\sim n \chi^3$. 

Second is the kinetic energy $K_{ij}$. We discretize the Laplacian on the grid using finite difference, i.e. 
$\partial^2\phi/\partial x^2 \approx [\phi(x+1/2^n)-2\phi(x)+\phi(x-1/2^n)]/2^{2n}$. 
In the quantics representation, this second derivative can be written explicitly as a ``Matrix Product Operator'' (MPO, an MPO is to matrices what MPS are to vectors) 
$\Delta_{\vec r_1\vec r_2}$ with a very small bond dimension $\chi=4$ \cite{kazeev2012}. 
The calculation of $K_{ij}$ is therefore put in the form
$K_{ij} = V_0\sum_{\vec r_1 \vec r_2} \Phi^i_{\vec r_1} \Delta_{\vec r_1\vec r_2} \Phi^j_{\vec r_2}$ which is a MPS-MPO-MPS product. We are back to the standard algorithms of the MPO/MPS toolbox.

Next comes the contribution from the nuclei potential $P_{ij}$. The functions
$1/|\vec r - \vec R_\alpha|$ are given to the TCI algorithm which produces an MPS $P_{\vec r}^\alpha$. Our numerical experiments show a fast convergence: the relative accuracy for $P_{\vec r}^\alpha$ is around $10^{-3}$ for $\chi = 60$ and  $10^{-5}$ for $\chi=100$. We arrive at $P_{ij} = V_0\sum_{\vec r, \alpha} \Phi^i_{\vec r} P_{\vec r}^\alpha \Phi^j_{\vec r}$ which is a direct extension of the MPS scalar product. It proves more efficient to compute this matrix element for every nucleus $\alpha$ and then sum the whole, because having one single MPS for all the potentials tends to require bigger $\chi$ for a given accuracy.

Last, we need to calculate the interaction matrix elements $V_{ijkl}$. The function
$1/|\vec r_1 - \vec r_2|$ is given to the TCI algorithm which produces an MPO $U_{\vec r_1,\vec r_2}$. Here as well, $\chi \approx 100$ is needed to reach a relative accuracy of $10^{-4}$. We arrive at $V_{ijkl} =V_0^2 \sum_{\vec r_1 \vec r_2} \Phi^i_{\vec r_1}\Phi^j_{\vec r_1} U_{\vec r_1\vec r_2} \Phi^k_{\vec r_2} \Phi^l_{\vec r_2}$. To evaluate these matrix elements we first form the element-wise product $\Phi^i_{\vec r}\Phi^j_{\vec r}$ for all pairs $ij$, then compress them using SVD into an MPS
$\Phi^{ij}_{\vec r}$. Then, we're back to the calculations of MPS-MPO-MPS products
$V_{ijkl} = V_0^2\sum_{\vec r_1 \vec r_2} \Phi^{ij}_{\vec r_1} U_{\vec r_1\vec r_2} \Phi^{kl}_{\vec r_2}$. This completes the suite of algorithms.

\subsection{Validation of the whole algorithmic chain} To validate the entire procedure, we make use of Gaussian orbitals for which all these contributions are known analytically. We use the package pyscf \cite{sun2018} for this benchmark.
Figure \ref{fig:2} shows a calculation of all the matrix elements for the LiH molecule in the STO-6G basis set (first three panels) together with the resulting CCSD(T) calculation (last panel). This is a mere validation that we can calculate the {\it inputs} of the many-body problem, to be solved by any suitable mean.  
We find that $n=16$ and $\chi = 17$ (with $\chi\approx 60$ for the MPS
$P_{\vec r}^\alpha$ and MPO $U_{\vec r_1,\vec r_2}$) are sufficient to reach chemical accuracy for all matrix elements as well as  the final total energy. 
Interestingly though, a slightly coarser discretization of $(2^{12})^3 \approx 6. \ 10^{10}$ points is not sufficient, showing that the arbitrary resolution available with tensorized orbitals is really needed. 

For a target accuracy at $1m$Ha on the energies ($n=16$, $\chi=17$), the computing time to
compute all matrix elements is CPU $\approx n [78 A(M) + 0.18 B(M)]$ milli-seconds on a single computing core with $A(M) = M(M+1)/2$ and $B(M)= M(M+1)(M^2+M+1)/8$ counting the number of different
elements in respectively $S_{ij}$ and $V_{ijkl}$. This translates into about $30$s for the present example which can be trivially parallelized and further optimized. 
While this is not competitive with the specific method for Gaussian orbitals, it is fast enough
for \emph{not} being the bottleneck of the calculation, in particular 
when an accurate many-body solver such as CCSD(T) [CPU$\sim M^7$ at half filling] is used.

\section{Natural orbitals as single tensorized orbitals}
Tensorized orbitals is really just a tool that enables one to use a much wider class
of orbitals that was possible before: it solves the Coulomb matrix integral bottleneck.
We surmise that this tool may be used in many different ways. An obvious application, that we leave to future work, is the construction of new basis sets for atoms, using e.g. Slater orbitals that should have better convergence properties than Gaussian in the tails and close to the nuclei. 
Another is to use tensorized orbitals to describe molecular or natural orbitals that we explore in this section. Our CCSD(T) benchmark for $H_2$ and $CH_4$ shows that a gain in accuracy of respectively a factor $4$ and $2$ can be obtained with respect to a double zeta calculation (cc-pvDz basis set) that uses the same number of orbitals. 
This gain can be further improved to a factor $10$ using a simple extrapolation scheme
and is also present in other simple molecules (see Appendix \ref{basis_set_error}). These preliminary results indicate that a path \emph{does} exist for an increase in accuracy of one order of magnitude at no additional computing cost.

\subsection{Basis set versus correlation errors}
Calculations performed with five zeta gaussian basis sets (e.g. cc-pv5z) can be considered as accurate while double zeta basis sets (e.g. cc-pvDz) suffer from strong inaccuracies \cite{ren2023}. The problem is that even for a moderately big molecule like benzene, it is difficult to go beyond double zeta while properly accounting for correlations \cite{eriksen2020}.
The question is therefore simple: can tensorized orbitals provide some of the five zeta accuracy while working with basis sets of the size of the double zeta?
Let us define the problem more precisely: one may work in a regime where $M_{AO}$ (the number of underlying atomic orbitals, say Gaussians and/or Slater) is much larger than the number of (molecular or natural) orbitals $M$ used in the many-body solver (the bottleneck of the calculation). We define the error $\epsilon(M,M_{AO}) = |E(M,M_{AO})-E(M=\infty,M_{AO}=\infty)| = \epsilon_{\rm BS}+ \epsilon_{\rm cor}$ as the sum of the "basis set" error $\epsilon_{\rm BS}(M,M_{AO}) = |E(M,M_{AO})-E(M,M_{AO}=\infty)|$ and the "correlation" error $\epsilon_{\rm cor}(M) = |E(M,M_{AO}=\infty)-E(M=\infty,M_{AO}=\infty)|$. Typical quantum chemistry calculations use $M_{AO}\sim M$. However a tensorized orbital can be constructed out of tens or hundreds of atomic orbital, it is still a unique orbital and performing calculations with it is essentially independent on how it has been constructed. The problem therefore consists of two parts: first, evaluate how large is the part of the error $\epsilon_{\rm BS}$ that can potentially be removed; second, find an algorithm to get rid of it.

\begin{figure}[h!]
    \centering
    \includegraphics[width=\linewidth]{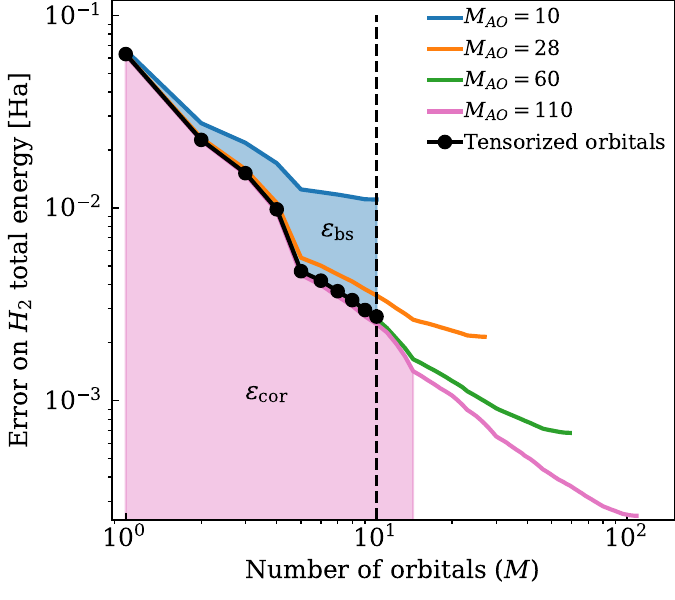}
    \caption{Error on the ground energy of the $H_2$ molecule for different basis sets. The calculation uses $M$ orbitals constructed optimally out of $M_{AO}$ atomic orbitals (colored lines) or using our enrichment algorithm (black circles, $n=20$). Pink shaded area: $\epsilon_{\rm cor}$; Blue shaded area: $\epsilon_{\rm BS}$. The interatomic distance is $d_{H-H} = 1.4013 ~ a_0$, the reference energy is $E_0 = -1.17447498 ~ {\rm Ha}$ \cite{kolos1968} and
    the many-body problem is treated exactly with \texttt{pyscf}. 
    \label{fig:enrichment}
    }
\end{figure}
\subsection{Application to the $H_2$ molecule} Let us estimate $\epsilon_{\rm BS}$ and $\epsilon_{\rm cor}$ for the $H_2$ molecule. 
We perform the following calculation:
(i) We perform a CCSD calculation (exact here since there are only 2 electrons) of $H_2$ in the cc-pvXz basis, X=D,T,Q,5 corresponding to $M_{AO}=10, 28, 60$ and $110$.
(ii) For each calculation, we diagonalize the one-body density matrix $\rho_{ij} = \langle c^\dagger_i c_j\rangle$. Its eigenstates are the 
natural orbitals of the problem \cite{davidson1972} and the eigenvalues the corresponding filling factors.
(iii) We perform a second series of CCSD calculation keeping only 
$M$ natural orbitals with the highest filling factor.
The results are shown in Fig. \ref{fig:enrichment} (thin lines). 
As one increases the ratio $M_{AO}/M\rightarrow \infty$, one observes the convergence towards a universal curve where the error is entirely controled by the dynamical correlations -- not by the choice or size of the basis set -- this is the continuum limit where $\epsilon = \epsilon_{\rm cor}$. Note that this continuum limit is \emph{not} the usual CBS limit, since the latter also implies taking $M\rightarrow \infty$. For $M=10$ the correlation error is $\epsilon_{\rm cor} = 2.5$ mHa, it is more than four times smaller than the full error $\epsilon = 11.1$ mHa of the cc-pvDz calculation that uses $M=M_{AO}=10$ (end of the blue line). 
So far, this calculation is contrieved because we performed an expensive CCSD calculation with $M=110$ in step (i). Nevertheless it indicates that the potential for a large gain in accuracy is present, we just need an algorithm to build the optimized orbitals directly. 
We have repeated the procedure (i)-(iii) for several molecules (see Appendix \ref{basis_set_error}) and consistently found a large value of $\epsilon_{\rm BS}$ of more than $100$mHa for the cc-pvDz basis. More importantly, the remaining error $\epsilon_{\rm cor}(M)$ depends only on $M$ (not on the precise and delicate construction of the basis set) and is therefore free of any arbitrariness in its construction. Altogether we consistently found that the final error with respect to CBS could be lowered by a factor ten or more.

\subsection{Enrichment algorithm} The preferred way to optimize the orbitals will probably be directly in real space independently from Gaussians. However, as a first attempt, we show here an enrichment algorithm that provides the optimum error for $M=10$ without paying the high price of step (i) (black circles in Fig. \ref{fig:enrichment}). The enrichment algorithm goes as follows: (a) We choose the $M^*=15$ atomic orbitals of lowest energy and remove them from the cc-pvQz basis set. (b) We solve the many-body problem with these orbitals. (c) We construct the corresponding natural orbitals. (d) We keep the $M<M^*$ natural orbital of highest filling factor; we enrich the basis set with $M^*-M$ new atomic orbitals taken from the remaining ones of the cc-pvQz basis set. (e) We repeat steps (b,c,d) until there are no orbitals left in the cc-pvQz basis set. While this algorithm is rather crude, we find that it provides the optimum error for $M=10$ while \emph{never} solving the many-body problem for more than $M^*\sim M$ orbitals. This is a first step towards getting rid entirely of the basis set error.

\begin{figure}[h!]
    \centering
    \includegraphics[width=\linewidth]{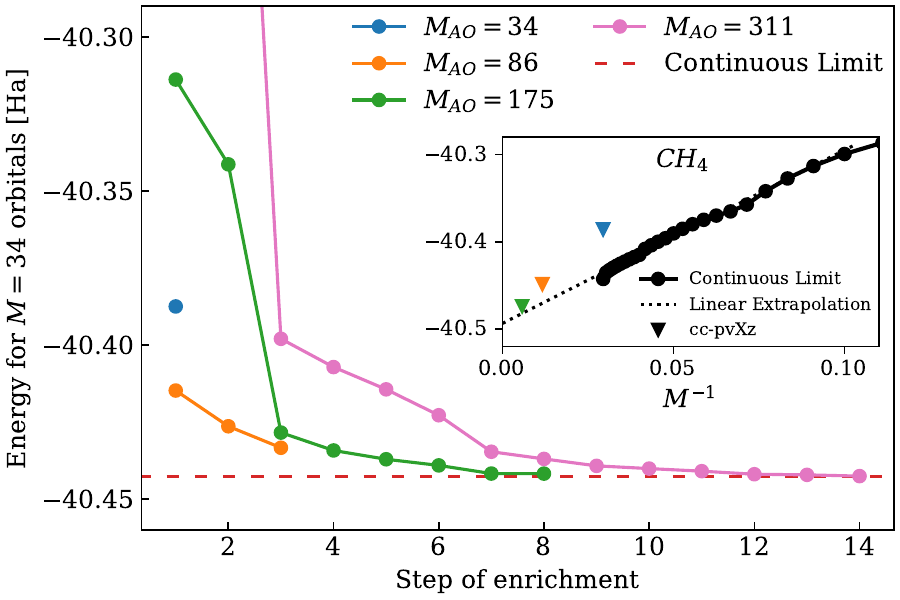}
    \caption{Results of enrichment process on the $CH_4$ molecule with $M=34$, $M^{\star} = 54$ versus number of enrichment steps. The enrichement is performed from the cc-pvDz ($M_{AO}=34$, blue, single data point), cc-pvTz ($M_{AO}=86$, orange), cc-pvQz ($M_{AO}=175$, green) and the cc-pv5z ($M_{AO}=311$, pink) basis set. Both green and pink curve reach the continuous limit.
    The inset shows the CCSD(T) calculations of the energy versus the number of optimum orbitals $M$ for $M\le 34$ computed by enrichment (black circles).
    Also shown are regular cc-pvXz CCSD(T) calculations (color triangles).
    Parameters of the $CH_4$ molecule: $d_{C-H} = 2.0541a_0$ ; $\theta_{CH_4} = 109.5^{\circ}$. The enrichment was done with $n=20$.
    \label{fig:enrichment_extrapolation_CH4}} 
\end{figure}

\subsection{Application to the $CH_4$ molecule} We repeat the same protocole on the $CH_4$ molecule using $M=34$ (size of the cc-pvDz basis set) and adding $M^*-M=20$ atomic orbitals at each step of the enrichment. The results are shown in Fig. \ref{fig:enrichment_extrapolation_CH4}. The main panel show the convergence with respect to the enrichement steps when enriching with orbital from the triple zeta cc-pvTz basis set (orange), cc-pvQz (green) and cc-pv5z (pink). We observe that the latter two converge to the same (continuum) limit. This is another example of construction of optimum orbitals that
get rid of the $\epsilon_{\rm BS}$ error entirely using the enrichment algorithm. The cost 
of the enrichment algorithm is relatively mild: assuming that one is dominated by the many-body solver that scales as CPU $\sim M^\alpha$, the total cost is CPU $\sim (M^*)^\alpha M_{AO}/(M^*-M)$. For the optimum value of $M^*$, one arrives at
CPU $\sim a(\alpha) M^\alpha (M_{AO}/M) $ with 
$a(\alpha)\equiv  [\alpha/(\alpha-1)]^{\alpha-1} \alpha$ ($a(\alpha)\sim e\alpha$ for large $\alpha$), i.e the increase of computing time is linear in $M_{AO}/M$.

The inset of Fig. \ref{fig:enrichment_extrapolation_CH4} shows the obtained continuum energy versus $1/M$ for $M\le 34$ i.e. for many-body calculations that use less orbitals than the double zeta calculation. Also shown are the standard results in the cc-pvXz basis sets (X=D,T,Q, colored triangles). 
Very interestingly, the continuum limit is intrinsic, i.e. it does not depend on the choice and construction of a basis set. It just depends on the number of orbitals used, so one
may surmise that the CBS limit is well behaved.
We find that a simple linear extrapolation of our data from $M\le 34$ provides
a good estimate of the CBS limit with an error $\sim 10$ mHa, a tenfold improvement with respect to the cc-pvDz calculation with the {\it same} value of $M$ (see Appendix \ref{basis_set_error}) for similar statements on a few other molecules).

\section{Constructing tensorized orbitals directly from real space calculations} Ultimately, we would like to contruct the optimized tensorized orbitals directly following e.g. algorithms in the spirit of \cite{Valeev2023}. We end this letter by taking a first step in this direction, calculating the (tensorized) ground state of the $H_2^+$ and $HeH^{2+}$ ions (one electron and two nuclei) directly using MPO/MPS techniques. This is an interesting case because chemical bonds are not easy to represent with high accuracy with Gaussian orbitals (see Appendix \ref{gauss_v_dmrg}). We seek a tensorized orbital $\Phi_{\vec r}$ that minimizes the energy of the $H_2^+$/$HeH^{2+}$ ion. Here, we take advantage of the fact that the Schr\"odinger equation for the wavefunction of the electron has already been put into MPO/MPS form, i.e. we are looking for the lowest eigenenergy of the one electron Hamiltonian, or equivalently the minimum of,
\begin{equation}
E = V_0 \min_{\Phi_{\vec r}} \sum_{\vec r_1,\vec r_2}
\Phi_{\vec r_1}[ \Delta_{\vec r_1,\vec r_2} + P_{\vec r_1}
\delta_{\vec r_1,\vec r_2}]\Phi_{\vec r_2} 
\label{eq:h2}
\end{equation}
where $\delta_{\vec r_1,\vec r_2}$ is the kronecker symbol. Performing such a minimization is exactly what the celebrated DMRG algorithm does, although usually in a totally different context (each index usually corresponds to the occupation of an orbital while here
they label the different scales of a one-particle problem). Hence we can rely on any existing implementation of DMRG to calculate the orbital (here the quimb package \cite{gray2018}). The results are shown in Table \ref{table:DMRG}. DMRG easily outperforms the precision that we could obtain 
with two hundred Gaussians \citep{jensen2013} by one order of magnitude. More importantly, the
result is a single orbital whose rank is not significantly higher than those of simple Gaussians or Slaters.

\begin{table*}[t!]
\setlength{\tabcolsep}{.7em}
\begin{tabular}{ |p{3 cm}||p{4cm}|p{4cm}|p{4.5cm}|}
\hline
 \multicolumn{4}{|c|}{Error on the ground state energy} \\
 \hline
\hline
 Method & \centering $H$ & \centering $H_2^+$ $(d_{H-H} = 2~ a_0)$ &  $HeH^{2+}$ $(d_{He-H} = 1.4634 ~ a_0)$ \\
 \hline
 \hline
 cc-pvDz  basis set  &  $+0.72 ~m{\mathrm{Ha}}$\hfill  \hfill [28 gaussians]    &  $+~2.37 ~m{\mathrm{Ha}}$ \hfill  \hfill [56 gaussians] & $+~ 7.17 ~m{\mathrm{Ha}}$ \hfill  \hfill [56 gaussians] \\
  \hline
 cc-pv5z  basis set &   $+5.46 ~\mu{\mathrm{Ha}}$  \hfill  \hfill[110 gaussians]    &  $+14.5~\mu {\mathrm{Ha}}$  \hfill  \hfill[220  gaussians] &  $ +64.6 ~\mu{\mathrm{Ha}}$ \hfill  \hfill[220 gaussians] \\
  \hline
 DMRG with $n=20$ &  $+1.45  ~\mu{\mathrm{Ha}}$  \hfill  \hfill [$\chi = 54$] &  $+1.22 ~\mu{\mathrm{Ha}}$ \hfill  \hfill[$\chi = 53$] &  $\equiv 0  ~\mu{\mathrm{Ha}}$\hfill  \hfill [$\chi = 58$] \\
  \hline \hline
 Reference energy    & $-~0.5 ~ \mathrm{Ha}$  &  $-~1.1026342144949 ~ \mathrm{Ha}$ \hfill \cite{kolos1968} &  $-~2.7136469(2) ~{\mathrm{Ha}}$ \\
 \hline
\end{tabular}
\caption{Single orbital calculation: linear combination of Gaussians versus DMRG.  Error on ground state energies of a single electron in different ions.  The cc-pvDz and cc-pv5z rows are obtained with \texttt{pyscf} simulations. The MPS are calculated with the DMRG algorithm. A $n=20$-MPS performs better than the most elaborate Gaussian basis set (cc-pv5z) by an order of magnitude. For $HeH^{2+}$, the energy obtained by DMRG was more precise that what we could find in the litterature, we use it as reference. 
\label{table:DMRG}}
\end{table*}

\section{Conclusions} Tensor network techniques were invented in the context of solving many-body problems, mostly in one dimension, and for a long time were mostly confined to this application. In the last few years important technical developments have appeared and these techniques are stepping out of their original context. In particular the TCI algorithm provides a natural bridge to map problems apparently unrelated to tensor networks onto the MPO/MPS toolbox. In this article we have shown that a combination of the quantics representation, TCI and the traditional MPO/MPS toolbox provides all the necessary ingredients to represent atomic, molecular or natural orbitals with very high - perhaps unprecedented - accuracy.
This methodology provides new possibilities for constructing accurate basis sets, lifting the strong constraint of having to express them solely in terms of Gaussians. Future work will include the integration of these tensorized orbitals, ideally within a standardized format, with the existing quantum chemistry packages
as well as the exploration of new ways to construct basis sets opened by this format.

\section*{Acknowledgment} We thank Miles Stoudenmire, Emanuel Gull and Steve White for interesting discussions. XW acknowledges funding from the Plan France 2030 ANR-22-PETQ-0007 “EPIQ”, from the French ANR DADDI, from the AI program of the French MESRI and from the CEA-FJZ AIDAS program.


\appendix
\section{Some pointers to the relevant MPO/MPS litterature.}\label{MPO_litt}
This article makes heavy use of several tensor network concepts that
are well known in some communities but new in others. Here, we gather a
few pointers to the litterature that may be useful for people new to the field.
\begin{itemize}
\item A concise and gentle introduction to tensor networks, in the context of quantum computing, may be found in the Appendix A of \cite{ayral2023}.
\item An introduction to the quantics representation of functions as tensors can be found in \cite{nunez2024}.
\item an introduction to the Tensor Cross Interpolation algorithm can be found in Section 3 of \cite{nunez2022}. Another introduction, together with an advanced presentation of the most advanced algorithms can be found in \cite{nunez2024}.
\item Most of the litterature on MPO/MPS algorithms focus on many-body solvers for one dimensional models of interacting spins or electrons.
Classic reviews include \cite{verstraete2008,schollwoeck2010,Bridgeman2017}. The algorithms presented in these references, such as DMRG, TEBD or TDVP can be directly used in the quantics representation although it might be interesting to build special versions optimized for quantics.
\item Interesting material can also be found in the documentation of tensor network open source libraries such as iTensor, TenPy (https://tenpy.readthedocs.io/en/latest/) and tensor4all
\cite{nunez2024}
\end{itemize}

\section{Controling the accuracy of TCI-built MPS.}\label{tci_acc}
\begin{figure}[h!]
    \centering
    \includegraphics[width=\linewidth]{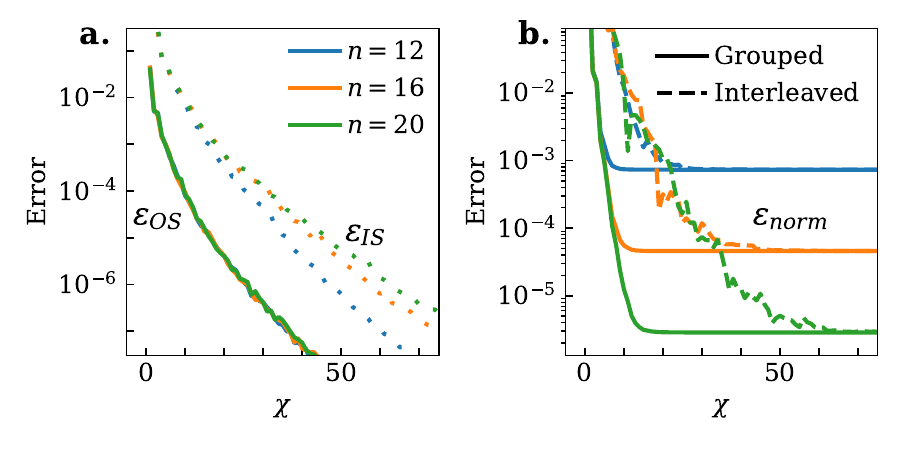}
    \caption{Errors of the MPS interpolating the 1s orbital of the Hydrogen atom versus its rank $\chi$. {\bf a.} In sample error ($\epsilon_{\mathrm{IS}}$) in dotted lines and out of sample error ($\epsilon_{\mathrm{OS}}$) in solid lines. Both are converging exponentially to $0$.
   {\bf b.} Error on the integral $\epsilon_{\mathrm norm}$. Here, the exact integral is  $\int \phi(x) \phi^{\dagger}(x) dx = 1$.
    \label{fig:errMH_v_errInt}}
\end{figure}
During the TCI construction of a tensorized orbital $\phi_{\rm MPS}(\vec r)$ from a known orbital $\phi(\vec r)$ (e.g. a Slater or a sum of Gaussians), we monitor three different types of error:
\begin{itemize}
\item The \emph{in-sample} error $\epsilon_{\mathrm IS}$ corresponds to the maximum error observed during the TCI construction between the actual function that is being tensorized and its tensorized approximation. We define it as,
\begin{equation}
\epsilon_{\mathrm IS}=\max_{\vec r_i\in Q} | \phi(\vec r_i) - \phi_{\rm MPS}(\vec r_i) |  
\end{equation}
where $Q$ is the set of observed points used by TCI in the interpolation process. In machine learning terminology, this would be called a training set error albeit a very conservative one since the algorithm consistently seeks points where this error is maximum.
\item The \emph{out-of-sample} error  $\epsilon_{\mathrm OS}$ corresponds to the average error on a set $P$ of $N_P$ randomly selected points ($N_P=500$ typically). In order to select these points in relevant regions (where e.g. the function is non vanishing), we obtain them by sampling 
$|\phi(\vec r)|^2$ using the Metopolis-Hastings algorithm. 
We define this error as :
\begin{equation}
\epsilon_{\mathrm OS}=\frac{1}{N_P} \sum_{\vec r_i \in P} | \phi(\vec r_i) - \phi_{\rm MPS}(\vec r_i) |
\end{equation} 
 In machine learning terminology, this error would correspond to the error associated with the test set.
\item The \emph{normalization error} $\epsilon_{\mathrm norm}$ (or more generally integral errors) is a global measure of the error. It is defined, for an orbital that is supposed to be normalized to unity, as
\begin{equation}
\epsilon_{\mathrm norm} = \left|V_0 \sum_{\vec r} |\phi_{\rm MPS}(\vec r)|^2 - 1\right|
\end{equation}
\end{itemize} 
We have similar definitions for other objects such as the Coulomb potential $1/|\vec r_1-\vec r_2|$ or other objects that need to be cast onto MPO/MPS. We observe that all these errors decrease exponentially with the rank $\chi$. In the case of the normalization error, the error eventually saturates to $\epsilon_{\mathrm norm} \sim 1/2^n$ when it becomes limited by the discretization. 

Fig. \ref{fig:errMH_v_errInt} shows these three errors for the case of the 1s orbital of the hydrogen atom for $n=12,16$ and $20$. All three error closely follow each other within a $O(1)$ multiplicative factor and display a very fast decay.

\begin{figure}[h]
    \centering
    \includegraphics[width=\linewidth]{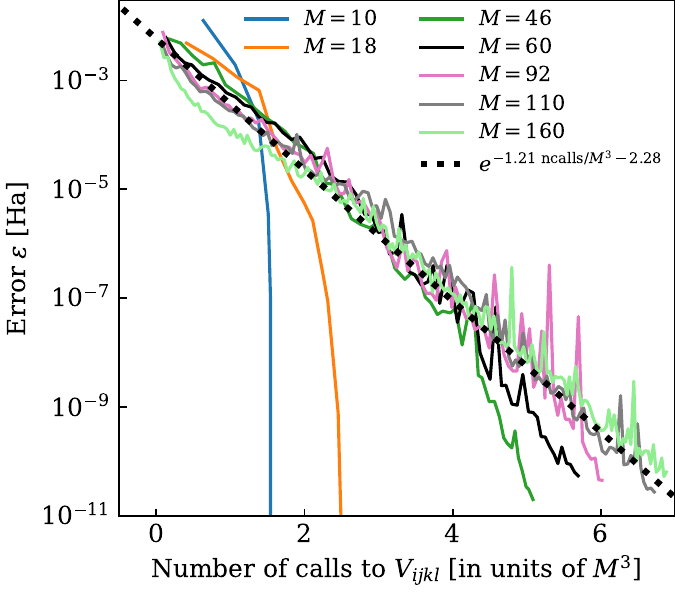}
    \caption{Error of the MPS interpolating the $V_{ijkl}$ tensor during its construction with the TCI algorithm. As the number of calls to the tensor grows, the error drops exponentially. The interpolation of the data is shown with the dotted line. The slope of the line is $-1.21$. The error is taken on samples of $10^4$ points taken randomly in the tensor.
    \label{fig:compV}}
\end{figure}

\begin{figure*}[t!]
    \centering
    \includegraphics[width=\linewidth]{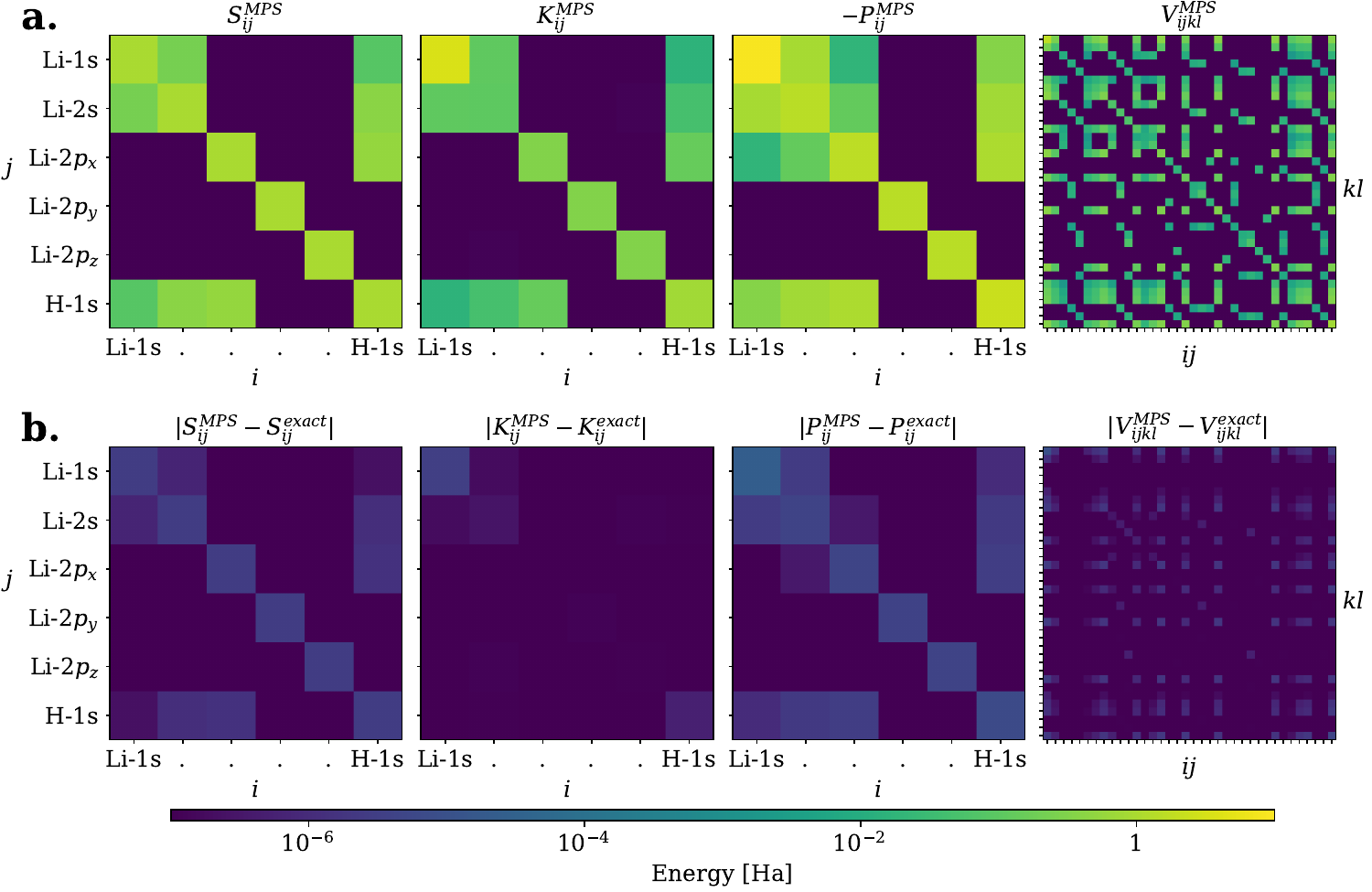}
    \caption{Upper panels {\bf a.}: Different matrix elements for the $LiH$ molecule in the STO-6g basis set. Lower panels {\bf b.}: corresponding error. This is the same data as for Fig. \ref{fig:2} of the main text in the $n=20$ case. From left to right: $S_{ij}$, 
 $K_{ij}$, $P_{ij}$ and $V_{ijkl}$.
    \label{fig:matelem_error}} 
\end{figure*}

\section{Direct compression of the $V_{ijkl}$ tensor using TCI.}\label{V_compression}
Large chemistry computations often require thousands of orbitals.
For such large values of $M$, storing the $V_{ijkl}$ tensor in memory can become a bottleneck, as it contains $\sim M^4$ elements. The computing part can also become a significant part of the computation, even when using Gaussians, in particular for relatively light methods such as DFT.
A common approach to alleviate this problem is to build a so-called "auxiliary basis" made of $O(M)$ Gaussians \cite{weigend2002}. The products 
$\phi_{ij}(\vec r) \equiv \phi_i(\vec r) \phi_j(\vec r)$ are expanded onto this auxiliary basis set (the associated tensor has $\sim M^3$ elements). Then, only the Coulomb matrix elements between the auxiliary orbitals need to be computed. This approach reduces the memory footprint to $\sim M^3$ and works well but it requires the fine tuning of the auxiliary basis which is a source of extra loss in accuracy.

Here, we use a different approach and compress $V_{ijkl}$ directly using TCI, therefore never building the full tensor. Note that this approach takes advantage of a unique feature of TCI with respect to other compressing approaches such as those based on the singular value decomposition: the algorithm does not need the entire tensor $V_{ijkl}$ to compress it, it only need to be able to compute \emph{some} of its elements.

Fig. \ref{fig:compV} shows our numerical experiment for a few test molecules.
We found that for a targeted error $\epsilon$ (Frobenius norm),
the number of $V_{ijkl}$ elements that need to be computed
scales as  $\sim (-\log\epsilon) M^3$, i.e. offers the same compression scaling as the auxiliary basis set approach. It has two distinct advantages however: its accuracy can be easily controlled and the method is fully automatic and agnostic with respect to the underlying basis set (or lack of).
This approach can be used to reduce the leading $O(M^4)$ contribution to the calculation of the
Coulomb matrix elements.

\section{Matrix elements of the $LiH$ Molecule} \label{matelems}
In the main text, we show the convergence of the matrix elements and the total energy (at the CCSD(T) level) for the $LiH$ molecule in the STO-6g basis set. The input of the many-body calculation are the integrals $S_{ij}^{\rm MPS}$, $K_{ij}^{\rm MPS}$, $P_{ij}^{\rm MPS}$,$V_{ijkl}^{\rm MPS}$  that need to be computed for the tensorized orbitals. Fig. \ref{fig:matelem_error} displays these matrices for the $n=20$ case (uper panels) along with the corresponding error on each matrix element (lower panels). This is the same data as Fig. \ref{fig:2}. We consistently find that that the error is more than 5 orders of magnitudes smaller than the value of the integrals.

\begin{figure}[ht]
    \centering
    \includegraphics[width=\linewidth]{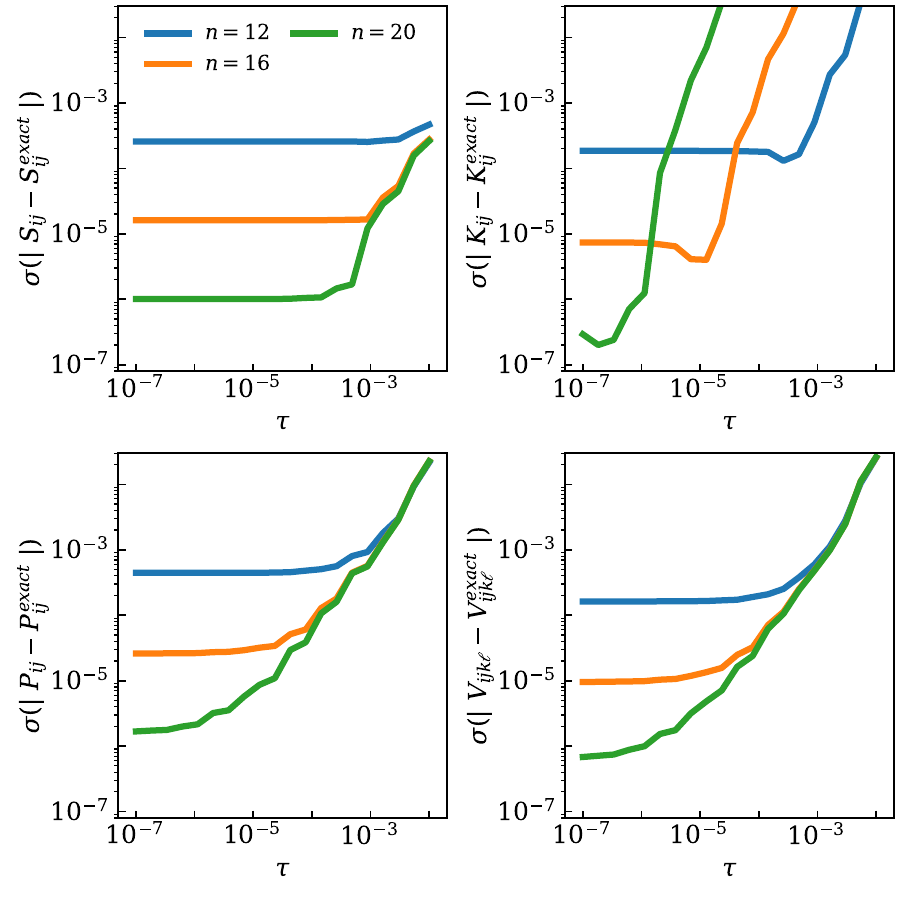}
    \caption{Error on the matrix elements for varying tolerance $\tau$. Here, the error is defined as the standard deviation of the distance between the target and the obtained result :  $\sigma(|S_{ij} - S^{\rm exact}_{ij}|)$. The matrix elements were computed for $H_2$ with the basis set cc-pvDz ($M=10$).
    \label{fig:scalings_accuracy}
    }
\end{figure}

\section{Accuracy of the matrix elements versus tolerance.}\label{matelem_acc_v_tol}
In the main text, we showcased the errors of the matrix elements versus rank $\chi$.
In practical calculations, we do not fix $\chi$ but rather a tolerance $\tau$ for the observed error requiring $\epsilon <\tau$ in all steps of the calculation. Indeed, different MPS can have different ranks for the same targeted tolerance, so this approach is much more robust.
Fig. \ref{fig:scalings_accuracy} shows the error of the different matrix elements versus $\tau$
for the $H_2$ molecule within the cc-pvDz basis.
We observe that, here, having $n=12$ and $\tau=10^{-3}$ is actually enough to obtain $1m$Ha accuracy on the matrix elements and eventually on the final energy. 

\begin{figure}[h]
    \centering
    \includegraphics[width=\linewidth]{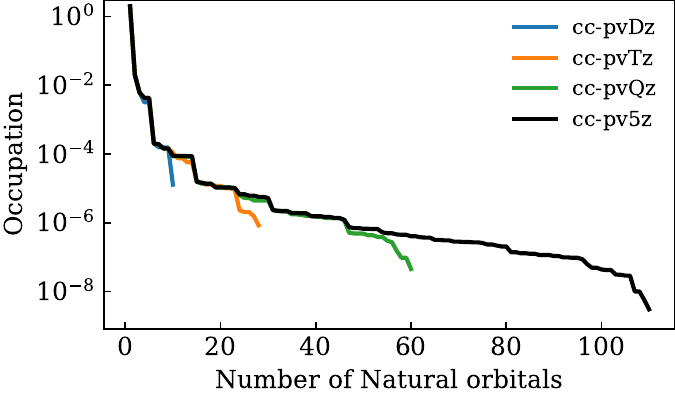}
    \caption{Occupation levels of each eigenstate of the 1-body density matrix $\rho_{ij} = \langle c_i^\dagger c_j  \rangle$. This is computed here for the $H_2$ molecule in different basis sets.
    \label{fig:nator_occupation}
    }
\end{figure}

\section{Natural Orbitals.}\label{natural_orbitals}
In Fig. \ref{fig:enrichment}, we have used natural orbitals to construct an optimum basis set of $M$ orbitals out of a much larger set of $M_{AO}$ (Gaussian) atomic orbitals. Introducing the
one-body density matrix $\rho_{ij} \equiv \langle c_i^\dagger c_j\rangle$, the natural orbitals
$\phi^N_\alpha(\vec r)$ are
obtained by digonalizing $\rho$,
\begin{equation}
\rho_{ij} = \sum_\alpha U^*_{\alpha i} \lambda_\alpha U_{\alpha j}
\end{equation} 
with
\begin{equation}
\phi^N_\alpha(\vec r) = \sum_i U_{\alpha i} \phi_i(\vec r)
\end{equation}
The eigenvalues $\lambda_\alpha$ are the occupation number of the orbitals:
$\lambda_\alpha = 1$ for a fully occupied orbital, $\lambda_\alpha = 0$ for an empty one
and $0<\lambda_\alpha<1$ for the interesting case of active orbitals responsible for electronic correlations. Fig. \ref{fig:nator_occupation} shows these occupation numbers for different basis sets of increasing size. We see that, as expected, only a few orbitals are occupied more than $10^{-3}$, but as we saw on Fig. \ref{fig:enrichment}, even lowly filled orbitals are necessary to obtain  accurate energies. We also observe a convergence towards the continuum limit as the size of the basis set is increased.

\section{Basis set error $\epsilon_{\rm BS}$ for a few selected molecules}\label{basis_set_error}

\begin{figure}[h!]
    \centering
    \includegraphics[width=\linewidth]{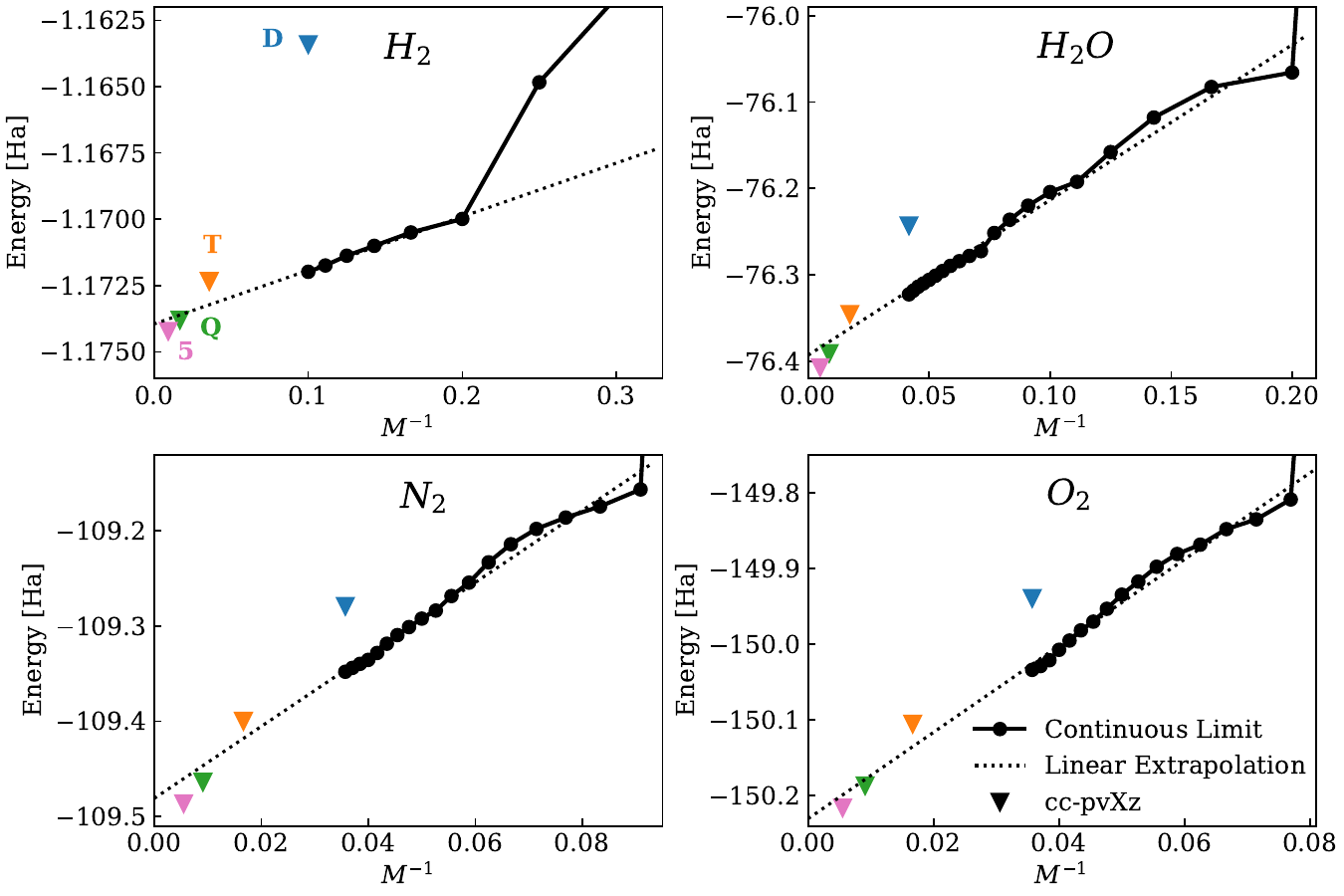}
    \caption{Couple cluster CCSD(T) calculation of the energy versus number of orbitals $M$ for a few molecules.
    Colored triangles: standard calculation using the cc-pvXz basis set with X = D (blue),T (orange), Q (green) and 5 (pink). Black circles: optimum orbitals in the continuum limit (constructed from natural orbitals, see text). Dotted lines: linear extrapolation from a simple linear regression.
   Molecular parameters: $d_{H-H} =1.40108a_0$ ; $d_{H-O} = 1.80869a_0$ ; $\theta_{H_2O} = 104.225^{\circ}$ ; $d_{N-N} = 2.07515a_0$ and $d_{O-O} = 2.270a_0$. 
    \label{fig:enrichment_extrapolation}} 
\end{figure}

In this section we try and assess the potential gain in accuracy that could be reached using tensorized orbitals. It is an extension of the discussion around Fig. 3 of the main text.

Fig.\ref{fig:enrichment_extrapolation} shows the energy versus number of orbitals $M$ for a
CCSD(T) calculation of four molecules ($H_2$, $H_2O$, $N_2$ and $O_2$). The triangles correspond to standard CCSD(T) calculations in the Gaussian basis set cc-pvXz with X = D (blue),T (orange), Q (green) and 5 (pink).
The black circle corresponds to the continuum limit obtained using the algorithm described in the main text
(essentially calculating the $M$ most important natural orbitals from a cc-pvQz calculation and 
checking that the results are converged with respect to the basis sets, i.e. unchanged when doing the same procedure with the cc-pv5z basis set). 

Two pieces of information can be extracted from this figure. First the last point of the continuum limit,
which corresponds to a calculation with the same value of $M$ as a cc-pvDz calculation is consistently more
accurate than the bare cc-pvDz result by a large value of $100mHa$ or more (except for $H_2$).  Second, the results in the continuum limit are intrinsic, i.e. they
depend only on $M$ not on how the basis set have been constructed. Hence the $M$-dependance of the energy
which is crucial for the extrapolation to the $M\rightarrow\infty$ CBS limit has its origin in the actual form of the correlations in the molecule, not in the ability of the scientist to construct a basis set that 
"exptrapolates well". Hence we conjecture that working in the continuum limit - which should be possible with tensorized orbitals - will lead to significant improvement on the accuracy of the CBS extrapolation.
To illustrate this point we have performed a linear extrapolation of our continuum limit (dotted line)
using a simple  linear regression. The results consistently provide an increase of precision of the order of a factor ten with respect to the cc-pvDz calculation (which corresponds to the largest value of $M$ considered in the continuum limit). For our $H_2$ benchmark where the exact energy is known, the calculation in the
cc-pvDZ basis set yields an error of $11.1$ mHa while the linear extrapolation to the CBS limit
has an error of $0.6$ mHa. Similarly for $H_2O$, the cc-pvDz error is
$150$ mHa while the linear extrapolation yields an error of $18$ mHa.

\section{Single orbitals: comparison between Gaussians and DMRG} \label{gauss_v_dmrg}
\begin{figure}[h]
    \centering
    \includegraphics[width=\linewidth]{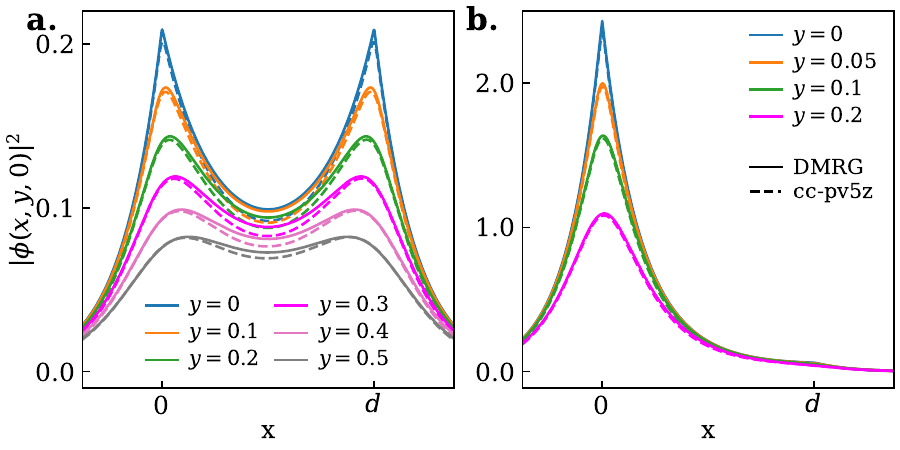}
    \caption{Profiles of density of probability $|\phi (x,y,z=0)|^2$ versus $x$ (fixed $y$) for the ground state of $H_2^+$ ({\bf a.}) and $HeH^{2+}$ ({\bf b.}). Solid line: DMRG result; dashed lines: best combination of Gaussian orbitals within the cc-pv5z basis set.  $d_{H-H} = 2~ a_0$  and $d_{He-H} = 1.4634 ~ a_0$. 
    \label{fig:repr_dmrg}
    }
\end{figure}

In Table \ref{table:DMRG} we showed that constructing orbitals directly with DMRG for ions yields very accurate energies, significantly better than what was obtained with Gaussians using very large basis set. Here, we examine the corresponding probability distributions and show that these
small differences in energy are associated to significant differences.
The results are shown in Fig. \ref{fig:repr_dmrg} for $H_2^+$ (left) and $HeH^{2+}$ (right).
The plot shows $|\phi (x,y,z)|^2$ versus $x$ for a fixed value of $y$ and $z=0$
($y=z=0$ corresponds to the nuclei-nuclei axis).  Full lines are converged DMRG result
(essentially exact) and dashed line are obtained from a linear combination of Gaussians with the
very accurate cc-pv5z basis set. 
For $H_2^+$, we observe that the combination of Gaussian displays significant deviations, especially close to the nuclei and in the tails (here in between the two protons), as expected. 
The discrepancy is harder to observe for $HeH^{2+}$ with the bare eye. However, the error on the  energy (a global probe) is larger in this case. 
Note also that there is considerable knowledge and work hidden in the construction of the Gaussian basis set while the DMRG approach is fully agnostic and automatic.

\section{Role of the bits ordering in the MPS rank.}\label{bit_ordering}
\begin{figure}[ht]
    \centering
    \includegraphics[width=\linewidth]{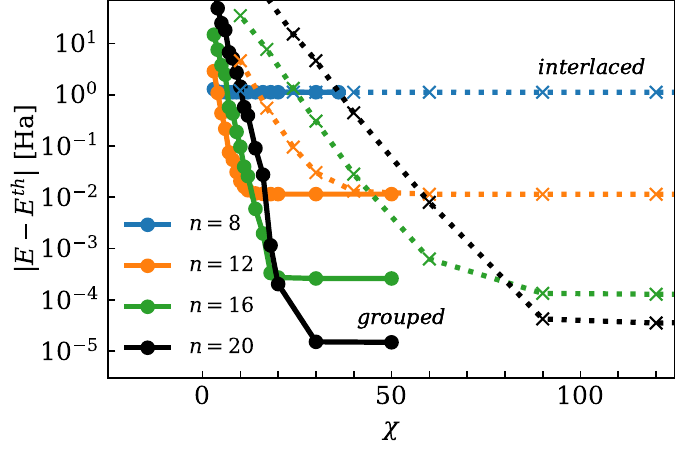}
    \caption{Error on the total energy of the LiH molecule in the STO-6g basis set versus bond dimension $\chi$ of the orbitals.
    This is a benchmark to show the accuracies of the MPS computed integrals compared to the exact gaussian based. The resolution is done at the CCSD(T) level.
    Distance between the nuclei: $2.8571 a_0$. All calculations with \emph{grouped ordering} are done with compression tolerance $\tau = 10^{-10}$ and for \emph{interleaved ordering} with $\tau = 10^{-5}$.
    The reference calculation uses the pyscf package \cite{sun2018}.
    \label{fig:ordering_LiH}}
\end{figure}

For the Quantics decomposition in 3D, several orderings of the bits are possible. We previously chose the \emph{grouped ordering} in which the indices are grouped by the coordinate they encode :
\begin{equation}
\small
\Phi_{x_1 ... x_n y_1  ... y_n z_1 ... z_n} = \prod_{a=1}^{n} M_{a}(x_a) \prod_{a=1}^{n} M_{a+n}(y_a) \prod_{a=1}^{n} M_{a+2n}(z_a)
\end{equation}
However, it is also possible to consider the \emph{interleaved ordering} : the indices are gathered by similar scales. It amounts to substituting Eq. \eqref{eq:mps_grouped} with :
\begin{equation}
\small
\Phi_{x_1y_1z_1x_2y_2z_2...x_ny_nz_n} = \prod_{a=1}^{n} M_{3a}(x_a) M_{3a+1}(y_a) M_{3a+2}(z_a)
\end{equation}

The choice of the encoding has an important impact on the bond dimensions of the MPS. For example, Gaussians are interpolated within numerical accuracy with $\chi = 11$ with \emph{grouped} ordering but for \emph{interleaved}, the same accuracy requires going above $\chi=150$. The laplacian MPO has $\chi=4$ in 3D with \emph{grouped ordering} but $\chi=7$ with \emph{interleaved} ordering. 
Overall, it seems that for the present application, \emph{grouped ordering} is significantly better. This is shown in Fig. \ref{fig:ordering_LiH} which reproduces the results of Fig.\ref{fig:2} for both orderings. 

\section{Multi-orbital MPS.}\label{oneBasis_MPS}

The different orbitals used in the calculations of this article are actually often very similar and only rotated, translated or dilated. For this reason, it would be efficient to combine all the orbitals in a single MPS. Such an MPS only requires an additionnal outer-leg to account for the different indices of the orbital. Therefore, we briefly consider the following encoding of all orbitals in a single MPS:
\begin{equation}
\small
\Phi_{\alpha \vec r} = M_0(\alpha)\prod_{a=1}^{n} M_{a}(x_a) \prod_{a=1}^{n} M_{a+n}(y_a) \prod_{a=1}^{n} M_{a+2n}(z_a)
\end{equation}
The computations of the matrix elements for such an MPS are greatly simplified: a single MPS-MPO-MPS contraction would yields the entire matrices $S_{ij}$, $H_{ij}$ and similarly for the $\Phi^{ij}_{\vec r}$ and $V_{ijkl}$ respectively. We leave such a program for future work and limit ourselves to investigating the impact on the rank $\chi$ of this scheme.

\begin{figure}
    \centering
    \includegraphics[width=\linewidth]{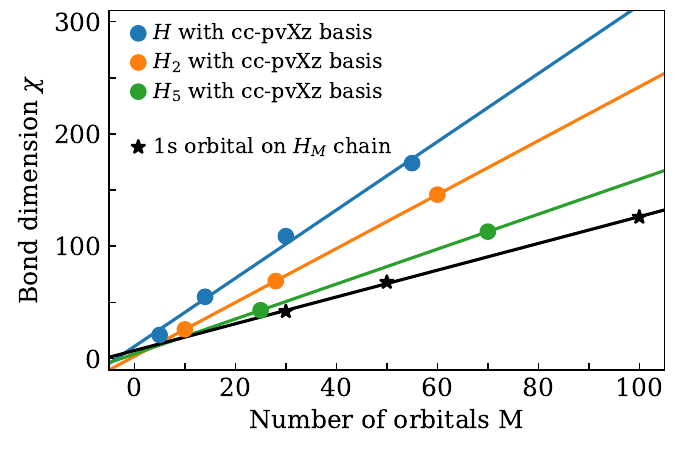}
    \caption{ Bond dimension of a single multi-orbital MPS for an accuracy of $\epsilon_{OS} = 10^{-4}$
    versus the total number of orbitals $M$ used to describe the molecular chain.
    We chose the standard cc-pvXz basis sets for chains of hydrogen ($H$, $H_2$
    and $H_5$, circles). From left to right X = D, T, Q, 5. We also considered the case of a single Slater orbital (1s) for a increasingly large chain $H_M$ (stars).  
    The symbols correspond to the data, the lines to linear interpolations.
    All calculations are done with $n=12$ and the step between atoms is $d = 1.4 a_0$. For chains longer than 20 atoms, we started a new line of atoms. 
    \label{fig:cost_multiorb}} 
\end{figure}

The results of a few numerical experiments are displayed in Fig. \ref{fig:cost_multiorb}. 
We observe that overall the bond dimension at given accuracy increases linearly with the number of orbitals $\chi\propto M$. 
We also observe that the more the basis sets are redundant
(e.g. correspond to simple translations/rotations of a few primitives), the smaller the slope. The storage of the orbitals using single-orbital tensorized orbitals scales as
$M$ against $M^2$ for the present scheme. Multi-orbital MPS therefore do not
seem to be advantageous memory wise. However the calculation of the Coulomb matrix elements
scales as $M^4$ for both single and multiple orbitals MPS, possibly with a significantly smaller prefactor for the latter. It follows that such a scheme could be advantageous when 
this computation becomes significant.

\bibliography{tensorized-orbitals}

\end{document}